\documentclass[aps,prl,twocolumn,showpacs]{revtex4}
\usepackage{epsfig}
\usepackage{graphicx}
\usepackage{amsmath}
\usepackage{times}
\bibstyle{apsrev.bib}

\newcommand{\beq}{\begin{equation}}
\newcommand{\eeq}{\end{equation}}

\begin{document}

\title{Polymer packaging and ejection in viral capsids: shape matters}

\author{I. Ali$^1$, D. Marenduzzo$^2$, J. M. Yeomans$^3$}
\affiliation{$^1$ Department of Physics, College of Science, PO Box 36,
Sultan Qaboos University, Al-Khodh 123, Oman \\
$^2$SUPA, School of Physics, University of Edinburgh, Mayfield Road,
Edinburgh EH3 9JZ, Scotland \\
$^3$Rudolf Peierls Centre for Theoretical Physics, 1 Keble
Road, Oxford OX1 3NP, England} 

\begin{abstract}

We use a mesoscale simulation approach to explore the 
impact of different capsid geometries on the packaging and 
ejection dynamics of polymers of different flexibility.  We find that
both packing and ejection times are faster for flexible polymers. For
such polymers a sphere packs more quickly and ejects more slowly than an
ellipsoid. For semiflexible polymers, however, the case relevant to
DNA, a sphere both packs and ejects more easily. 
We interpret our results by considering both the 
thermodynamics and the relaxational dynamics of the polymers.  The 
predictions could be tested with bio-mimetic
experiments with synthetic polymers inside artificial vesicles.
Our results suggest that phages may have evolved to be roughly
spherical in shape to optimise the speed of genome ejection,
which is the first stage in infection.
\end{abstract}

\maketitle

In this paper we study the packaging and ejection of polymers of
different flexibility into, and from, spherical and ellipsoidal
capsid shells. This is a model system for bacteriophages which
consist of a semiflexible polymer DNA (the genome) packaged into a
rigid container (the phage capsid) \cite{alberts}. This system
has recently attracted considerable theoretical attention
\cite{purohit3,kindt,purohit1,odijk,harvey,dewit,evilevitch,muthukumar,gabashvili}.
Here we use a numerical approach which was developed in \cite{ourJCP}
where it was shown to reproduce the pauses during packing which have
been observed experimentally \cite{smith}. Our main focus in the
present work is on DNA ejection and on the impact of different capsid
geometries on DNA packing and release. These issues had remained
virtually unexplored via simulations until now.


The packaged DNA is subject to strong energetic and entropic penalties
because it is contained within a capsid whose dimensions are typically
smaller than the DNA persistence length $\sim$ 50 nm \cite{smith,simpson}.
This builds up an enormous internal pressure $\sim$ tens of
atmospheres which the viruses or bacteriophages
exploit to provide the simplest of attack strategies. Typically
bacteriophages land on the surface of a bacterium and eject their
genome into the host cytoplasm simply by taking advantage of the
internal pressure which pushes the DNA out of the phage once the
capsid is opened.

The diversity in the naturally occurring shapes of viral capsids 
is remarkable \cite{baker,twarock}.
The shells of the phage DNA which infects
prokaryots, like E. coli or B. subtilis, are spherical or
quasi-spherical, stiff shells. 
For example, in the $\phi$29 phage, the 
DNA is fed into a 54 by 42 nm
icosahedral capsid \cite{tao}. 
On the other hand, viruses infecting higher
eukaryots, which rely on a more complicated infection strategy than
simple pressure-driven ejection, often have strikingly different,
much more elongated shapes. 
{An example here is the influenza virus which may be
e.g. $\sim$ 250 nm long and $\sim$ 100 nm wide \cite{ictv}.}
Moreover, data on the internal volume of capsids, although sketchy,
suggest that common spherical
phages like T7, $\lambda$ and HK97 pack their genome at a density
which is $\sim 10-20$ $\%$ larger than that encountered for 
slightly aspherical phages like $\phi$29 or T4 
(see Table 2 in Ref. \cite{purohit3}).
Our results suggest a possible explanation for these observations.

Work in this area is particularly timely because in vitro single
molecule experiments have led to significant quantitative insights on
the dynamics of packaging and ejection in vivo. Smith et al have
measured the rate of packaging and the force of the motor for the
$\phi$29 bacteriophage \cite{smith}, while more recent experiments
have characterized DNA ejection from the T5 and $\lambda$ capsids
\cite{experiment1,experiment2}. Both the
ejection and packaging rates have been shown to vary consistently {and
reproducibly} during the various stages of these processes.  

On the theoretical side, the effects of genome stiffness, 
excluded volume and electrostatics on 
the DNA packaging process have been investigated by 
thermodynamic theories and simulations 
\cite{purohit3,purohit1,odijk,kindt,davide,harvey,dewit,ourJCP}.
DNA ejection has also recently attracted a lot of attention
among theorists: in particular
the roles of the buffer in DNA ejection experiments in vitro 
\cite{evilevitch}, the relation of ejection to translocation
\cite{muthukumar} and ratchet \cite{phillips4} models  
have been the subject of recent studies. { Earlier
work \cite{gabashvili} pointed out that quasi-static 
analytic theories for DNA release require an assumption for the 
underlying main mechanism leading to friction during ejection.}

In this paper we use the stochastic rotation dynamics simulation model
\cite{mk} to compare the way in which flexible and  semiflexible 
polymers are packed into, and ejected from, spherical and ellipsoidal capsids 
of the same internal volume.  { Novel to this work
are the explicit simulations of the ejection kinetics, which correctly
capture non-equilibrium effects, and the focus
on the impact of capsid geometry on the physics of DNA packing and
releasing.} We find that the slower relaxation times of the
semiflexible chains leads to slower packing and ejection
rates.
A flexible polymer is ejected more quickly from an ellipsoidal
shell than a spherical one. However, at first sight surprisingly,
this situation is reversed for a semiflexible chain, which is ejected
more quickly from a spherical shell. We argue that this is 
a consequence of balances between the thermodynamic force driving
ejection and the ease with which the polymer can come to equilibrium 
within the confined space in the capsid. Recent advances in 
single molecule micromanipulation techniques \cite{smith} and in
DNA ejection imaging and analysis \cite{experiment1,experiment2}
put an experimental verification of these predictions within reach.
      
The polymer is a coarse-grained chain of $N=100$ beads  
joined by FENE springs, interacting via a potential $V$, 
\begin{equation}
4\epsilon \Sigma_{i} \left({\sigma}/{\mid \vec{r}_i
-\vec{r}_{i-1} \mid}\right)^{12}+\kappa\Sigma_{i} 
(\vec{r}_{i+1}-\vec{r}_{i})\cdot
(\vec{r}_{i}-\vec{r}_{i-1})
\label{potential}
\end{equation}
where $\vec{r_i}$ is the position of the $i^{th}$ bead. The first term is
the repulsive part of a Lennard-Jones potential which generates
excluded volume interactions between the beads. This is in tune with the
experiments where repulsive interactions dominate. The potential
parameters used were $\epsilon=k_BT$ and $\sigma =$2.5nm. $\kappa$ in
the second term in eq.~(\ref{potential}) is a bending rigidity
which sets the persistence length $l \sim \kappa \sigma / k_B T$. 
Here we set $l=0$ for a flexible polymer and $l=10 \sigma$ for
a semiflexible polymer. 
(We use 10 $\sigma$ to compromise between reaching typical genomic
stiffness -- 20 $\sigma$ under physiological conditions \cite{smith}
-- and feasible length and time scales in the simulations.)
The updating of the beads' positions and velocities is performed using
the velocity-Verlet molecular dynamics algorithm.

\begin{figure}
\begin{center}
\includegraphics[width=5.2cm]{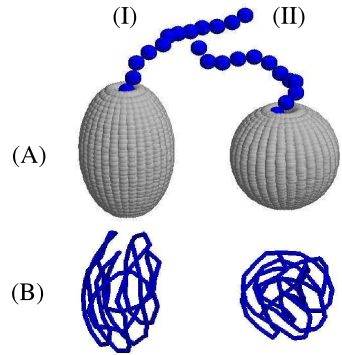}
\end{center}
\caption{Schematic representation of the simulation. A 
semiflexible polymer is
first packed into and then ejected from a rigid capsid. The
first row (A) shows the capsid and the dangling tail of the polymer
close to the end of a packing run. The second row (B) 
shows the configuration of the polymer chain inside the capsid.
Column (I) refers to a spherical capsid, and column (II) to an
ellipsoidal one. 
\label{capsid_shapes} }
\end{figure}

The capsid shapes, illustrated in Figure~\ref{capsid_shapes}, are
described by
\begin{equation}
f \equiv 1-\left[(x/a)^2+(y/b)^2+(z/c)^2\right]=0.
\label{ellipse}
\end{equation}
We choose $a=b=c=3.02 \sigma$ to model a sphere and $a=b=2.6 \sigma$
and $c=4.07 \sigma$ for an ellipsoid.  Each capsid is modeled as a
hard shell with a hole that permits the entrance of one bead at a
time. A repulsive force 
$k_B T/(\sigma f^4)$ is applied to
any bead which is at a point for which $\mid f \mid \le f_0 $,
where $f_0=0.2$ is a threshold. 
Our choices of 
$a$, $b$, $c$, and of $f_0$ lead
to the same volume 
available to the chain for both shapes, { which corresponds to
a packing fraction of $0.4$, comparable with previous numerical
work and typical phage densities \cite{kindt,purohit1}.
Qualitatively similar results to
the ones reported below have been found with $N=80$ and $N=120$ with
these capsid geometries, and with $N=200$ and a
packing fraction of $0.4$. }

The motor that feeds the polymer into the capsid is, in reality,
extremely complex \cite{simpson}. Here we use a simple model aimed at 
capturing the
basic physics. Essentially the motor has to (1) capture a bead and (2)
feed it into the capsid. This is accomplished by requiring the motor
to apply a radial force (of magnitude 5 $k_BT/\sigma$)
if the bead enters a cylinder of radius 
$0.7 \sigma$ and length $\sigma$  with origin at the capsid entrance.
The details {of this} mechanism do not affect our results.
Once captured, the bead is packed by applying a constant force towards
the centre of the capsid.  
Our simulations allow us to identify the minimum motor force which
is needed to achieve full packing.
To estimate this, we ran a set of packing simulations at different
motor force, and picked the lowest value of {the force} which still, on
average, packed the whole chain. Flexible and 
semiflexible polymers are respectively found to require a minimum motor 
force of 16 and 20 $k_BT/\sigma$ to be packed into a sphere, 
and of 18 and 26 $k_BT/\sigma$ to be packed into an ellipsoid.
{ In general the difference between the forces corresponding to the
sphere and the ellipsoid increases with packing fraction.}

The polymer is coupled to a coarse-grained solvent model, stochastic
rotation dynamics. This acts as a hydrodynamic thermostat allowing
momentum transfer between beads and allowing flows to be set up in the
surrounding fluid as a consequence of the bead motion. The solvent has
a viscosity $\sim$ 5 cP, comparable to that of cytosol.
The capsid is permeable to the solvent, which is the
physical situation for phage capsids. (We measure force and
time in simulation units in Figures 2--4. One time and force
simulation unit corresponds to 3 ns and 1.64 pN respectively.)

The polymer is initially configured randomly except for the
requirement that the first bead lies within the capsid and the rest
outside. The polymer is then equilibrated in this position before
opening the bead entrance and applying the feeding force. A single
bead is left out to initiate ejection once the motor
force is set to zero.  This is done after leaving time for the polymer to
equilibrate within the capsid.

Our simulations allow us to compare packing and ejection, a flexible
and semiflexible polymer, and a spherical and ellipsoidal
capsid. Figure~\ref{capsid_shapes} illustrates typical packed
configurations for the semiflexible polymer -- 
the polymers are ordered in spool-like domains { (although not in
an ideal inverse spool)} as predicted theoretically \cite{kindt,harvey}.

\begin{figure}
\begin{center}
\includegraphics[width=7.5cm]{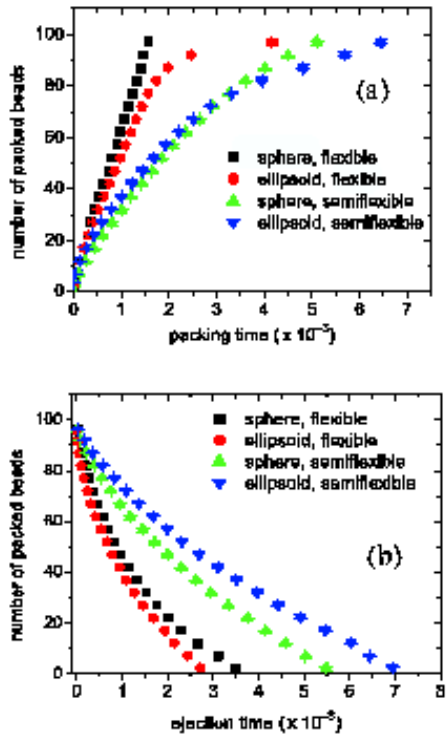}
\end{center}
\caption{Number of packed beads versus time
during (a) packaging and (b) ejection
for flexible and semiflexible polymers comparing a 
spherical and an ellipsoidal capsid.
\label{number_packed_beads}}
\end{figure}

Figure~\ref{number_packed_beads} shows the number of packed beads as a
function of time for both packing and ejection for the different chain 
flexibilities and capsid shapes. The motor force during packing was 
$26$ $k_BT/\sigma$ -- the minimum force to pack the 
semiflexible polymer into the ellipsoid, 
the case for which packing is hardest.
The most immediately striking feature is that packing and 
ejection times are considerably faster for flexible polymers. 
This is because relaxation times increase as the polymer becomes stiffer 
and e.g.\ once one bead has escaped it takes longer for the semiflexible chain 
to readjust itself so that a subsequent bead is in a position to escape.

Perhaps more surprisingly for flexible polymers the sphere ejects more
slowly than the ellipsoid whereas for semiflexible polymers 
the reverse is true.
The packing, on the other hand, proceeds more easily (it requires
a smaller minimum force) and more quickly (at equal packaging force
it takes less time) into the sphere whatever the flexibility of the polymer.
  
These results can be explained by considering a combination of
entropic and dynamic arguments. Both flexible and semiflexible polymers 
lose more entropy when they are  packed in an ellipsoid than when they are
packed in a sphere. Therefore, based on entropic  arguments alone, we would
expect the polymers to pack more easily in a sphere and to be ejected
more easily from an ellipsoid. This is the case for the flexible
polymers \cite{lattice}.  

For the semiflexible polymers, however, 
the sphere is faster  both in packing and
in ejection.
We believe this to be primarily a consequence of two effects. 
Firstly the beads of the semiflexible polymer suffer more from the constraint 
of being in the ellipsoid when they try to rearrange themselves as beads are 
ejected. Secondly the bending energy lost in packing is larger for the case of 
a sphere. These effects win over the entropic arguments and for the 
semiflexible polymer there is quite a pronounced advantage in ejection time 
for the sphere \cite{theory}. { The difference in ejection times between 
the sphere and the ellipsoid increases with the 
aspect ratio of the ellipsoid.}

\begin{figure}
\begin{center}
\includegraphics[width=7.5cm]{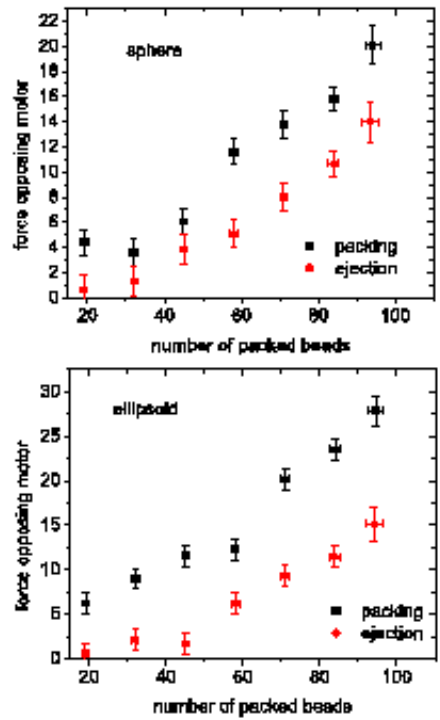}
\end{center}
\caption{Force opposing the motor { (in units of
$k_BT/\sigma$)} during packaging 
and ejection for a semiflexible polymer in (top) a spherical and (bottom) an
ellipsoidal capsid. 
\label{force_curves}}
\end{figure}

To investigate these non-equilibrium effects further we
plot, in Figure \ref{force_curves}, the force opposing the motor
during the packaging and ejection of a semiflexible polymer
for a sphere and an ellipsoid. We define 
such a force as the one felt by the bead inside the capsid which
is closest to the motor 
at a given time. It includes the force
due to local bending at the capsid entrance, the elastic
force due to the springs acting on the bead under observation, and the 
overall Lennard-Jones repulsion of the other beads, both in the
capsid and in the tail outside (i.e.\ we do not include the capsid 
contribution).

For both the capsid shapes there is {\it hysteresis}, i.e. the force
during the packing is larger than the one felt during ejection. This
shows that a significant portion of the resistance the motor
has to overcome during packing is due to dynamic dissipative
effects. The hysteresis is larger for the ellipsoidal capsid,
supporting the assertion that chain rearrangements are more difficult
in this case 
presumably due to the narrower
space close to the capsid tip.  { That dynamic effects are 
important in our simulations can also be appreciated by noting
that a set of simulations considering ejection from an ellipsoidal capsid
with a hole on the long side yielded an ejection time comparable to that
found from the sphere.}

\begin{figure}
\begin{center}
\includegraphics[width=7.5cm]{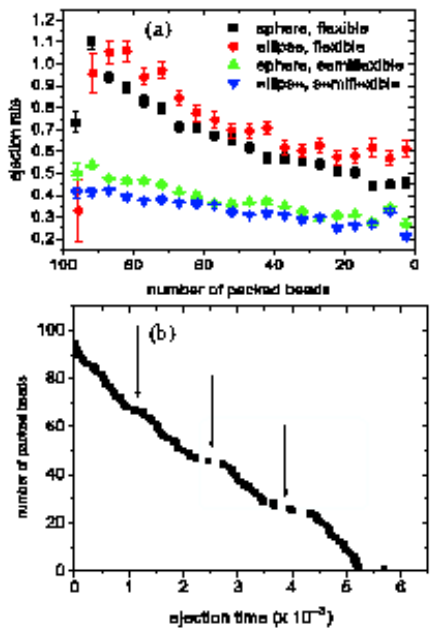}
\end{center}
\caption{(a) Ejection rate as a function of the number of packed beads 
for flexible and semiflexible polymers comparing a 
spherical and an ellipsoidal capsid. (b) Data from one
individual simulation of the ejection of a semiflexible polymer
from a sphere. Arrows indicate pauses { (corresponding to
rearrangements of the polymer inside the capsid)}.} 
\label{ejection_rate}
\end{figure}

Typically experiments report packing 
rates as a function of the number of packed beads \cite{smith}
and we therefore present similar data for ejection in Figure
\ref{ejection_rate}a. 
The ejection rate decreases as the number of
packed beads decreases for the flexible chain. This is because the 
force driving the ejection is the entropic penalty of confinement 
which decreases with decreasing packing fraction.  
For the semiflexible chain,
the rate decreases appreciably less quickly suggesting that entropic
considerations are less dominant. One can speculate that
the decreasing entropic force is offset by easier rearrangements
within the capsid as it empties.

The data in Figures \ref{number_packed_beads}b and 
\ref{ejection_rate}a are averaged over many
runs and thus the curves appear continuous. However, individual
runs indicate that, just as for packing, there are pauses in ejection
as the polymer rearranges itself within the capsid 
(Figure \ref{ejection_rate}b). 

In summary, we have compared the packing and ejection of flexible and
semiflexible polymers within phage capsids of variable geometry. The
behaviour is influenced not only by thermodynamic considerations but also by 
the relaxation time
of the polymers as they try to rearrange themselves within the
confined capsid geometry. For semiflexible polymers 
we find that a spherical shape leads to fastest ejection. One might
speculate that this is { one of the reasons why} most phages have 
evolved to be roughly spherical in shape. { We note
however that other phages, not relying on
pressure to eject their genome \cite{molineaux}, are
still spherical, thus suggesting that the mechanics and 
energetics of the proteins making up a viral capsid are
equally important in ultimately determining shape}. 
{ We also found that} non-equilibrium effects hamper
packing into an ellipsoidal phage, 
which may explain why aspherical capsids contain less DNA.
Our simulations suggest a series of single molecule bio-mimetic
experiments in which the dynamics of polymers of variable flexibility
undergoing packing-ejection cycles into and out of vescicles of
controlled shape are studied. 

{This work was supported by  EPSRC grant no. GR/R83712/01. 
We thank R. Golestanian for useful discussions.}

\end{document}